\documentclass[aps,prl,11pt,a4paper,twocolumn,groupedaddress,preprintnumbers,nofootinbib,floatfix]{revtex4-1}
\pdfoutput=1

\usepackage[top=2.9cm, bottom=2.1cm, left=2cm, right=2cm]{geometry}
\usepackage{comment}
\usepackage{graphicx}
\usepackage{url}
\usepackage[bookmarks, pagebackref=false]{hyperref}
\usepackage[usenames,dvipsnames]{xcolor}
\usepackage{dcolumn}
\usepackage{bm}
\usepackage{bbm}
\usepackage{amsmath,amssymb,amsfonts}
\usepackage{color}
\usepackage{hyperref}
\usepackage[Symbolsmallscale]{upgreek}
\usepackage{amsmath}
\usepackage{amsfonts}
\usepackage{amssymb,dsfont}
\usepackage[vcentermath]{youngtab}
\usepackage[all]{xy}
\usepackage{pstricks}
\usepackage{dsfont}%
\usepackage{cases}
\usepackage{placeins}
\usepackage{xspace}
\usepackage{cancel} 
\usepackage{slashed}
\usepackage[caption=false, labelformat=simple, listofformat=subsimple, labelfont=default, margin=5pt,
    justification=raggedright]{subfig}	

\newcommand{\beq}{\begin{eqnarray}}
\newcommand{\eeq}{\end{eqnarray}}

\newcommand{\bmp}{\noindent\begin{minipage}{16cm}}
\newcommand{\emp}{\end{minipage}\vskip 7mm} 

    \newcommand{\ii}{\mathrm{i}}

    \newcommand{\vw}{v_{\mathrm{w}}}

    \newcommand{\SU}{\mathrm{SU}}
    \newcommand{\SUR}{\mathrm{SU}(2)_{\mathrm{R}}}
    \newcommand{\SUL}{\mathrm{SU}(2)_{\mathrm{L}}}
    \newcommand{\Sp}{\mathrm{Sp}}

\def\lsim{\mathrel{\rlap{\lower4pt\hbox{\hskip1pt$\sim$}}
    \raise1pt\hbox{$<$}}}                
\def\gsim{\mathrel{\rlap{\lower4pt\hbox{\hskip1pt$\sim$}}
    \raise1pt\hbox{$>$}}}                

\begin{document}

\title{Radiatively Induced Fermi Scale in Grand Unified Theories 
}
\author{Tommi {\sc Alanne}}
\email{alanne@cp3.sdu.dk}
\affiliation{{CP}$^{ \bf 3}${-Origins} \& the Danish Institute for Advanced Study {\rm{Danish IAS}},  University of Southern Denmark, Campusvej 55, DK-5230 Odense M, Denmark.}
\author{Aurora {\sc Meroni}}
\email{meroni@cp3.sdu.dk}
\affiliation{{CP}$^{ \bf 3}${-Origins} \& the Danish Institute for Advanced Study {\rm{Danish IAS}},  University of Southern Denmark, Campusvej 55, DK-5230 Odense M, Denmark.}
\author{Francesco {\sc Sannino}}
\email{sannino@cp3.dias.sdu.dk}
\affiliation{{CP}$^{ \bf 3}${-Origins} \& the Danish Institute for Advanced Study {\rm{Danish IAS}},  University of Southern Denmark, Campusvej 55, DK-5230 Odense M, Denmark.}
\author{Kimmo {\sc Tuominen}}
\email{kimmo.i.tuominen@helsinki.fi}
\affiliation{Department of Physics, University of Helsinki, 
\& Helsinki Institute of Physics, \\
                      P.O.Box 64, FI-00014 University of Helsinki, Finland}
\begin{abstract}
We consider Grand Unified Theories in which the hierarchy between the unification and the Fermi scale emerges radiatively.
Within the Pati--Salam framework, we show that it is possible to construct a viable model where the Higgs is an elementary pseudo-Goldstone boson, and the correct hierarchy is generated. \\

{\footnotesize  \it Preprint: CP$^3$-Origins-2015-041 DNRF90 \& DIAS-2015-41 \& HIP-2015-35/TH}
\end{abstract}
\maketitle
\newpage

\section{Introduction}

The unification paradigm has motivated several extensions of the Standard Model (SM). Two time-honoured Grand Unified Theory (GUT) schemes stand out: the Georgi--Glashow  \cite{Georgi:1974sy}  and the Pati--Salam \cite{Pati:1974yy}. 
The former minimally unifies colour and electroweak (EW) symmetries in a higher-rank gauge group such as $\SU(5)$ or SO(10) \cite{Georgi:1975,Fritzsch:1974nn}
and predicts gauge-mediated proton decay.  The current 
  lower bound on the proton lifetime, set by SuperKamiokande to $\tau> 10^{34}$~y \cite{Nishino:2012bnw}, translates into a lower bound on the unification scale of the order of $\Lambda_{\mathrm{GUT}}\gtrsim 10^{15}$~GeV, which is also the natural scale for the realisation of type-I see-saw models \cite{Minkowski:1977sc,Yanagida:1979as,  Ramond:1979py, GellMann:1980vs,Mohapatra:1979ia}.  In the Pati--Salam model one unifies quarks and leptons by promoting the lepton number to the fourth colour. In this scheme the proton does not decay via gauge interactions, and therefore the previous bound does not apply.  On the other hand, spin-one leptoquarks would mediate  the $K_{\mathrm{L}}\rightarrow\mu^{\pm} e^{\mp}$ decay that is severely constrained by experiments leading to a lower bound~\cite{Parida:2014dba} $M>1.5\cdot 10^6$~GeV on their masses. Consequently, the lower bound on the Pati--Salam unification is $\Lambda_{\mathrm{PS}}>1.9\cdot 10^6$~GeV.

From experiments it is, therefore, clear that there are several well-separated energy scales involved in GUTs. At the very least, one needs the Fermi scale (where the EW symmetry breaks) and the scale where the theory unifies. These two scales are typically modelled via ad-hoc scalar sectors. 

It would seem appealing to us if there were only one common scale for all the scalar sectors, with the Fermi scale emerging radiatively.  We will show that such a scenario arises when the Higgs is an elementary pseudo-Goldstone boson (pGB) \cite{Alanne:2014kea}.  This opens the way to alternative scalar sector constructions in GUTs that we believe to be  more natural than traditional ones. 
  
 \section{A minimal Pati--Salam setup}
   
 We start with the Elementary Goldstone Higgs (EGH) scenario introduced in \cite{Alanne:2014kea, Gertov:2015xma} according to which the Higgs doublet lives in the $\SU(4)/\Sp(4)$ coset, and the EW symmetry,  $\SUL\times$U$(1)_Y$, is embedded in $\SU(4)$. 
 
 In comparison with the fundamental composite (Goldstone) Higgs idea \cite{Kaplan:1983fs, Kaplan:1983sm, Cacciapaglia:2014uja}, the major differences are:  the elementary case is amenable to perturbation theory; it is straightforward to endow the SM fermions with mass terms; it is possible to consider GUT extensions. 
 
As  minimal GUT extension of the EGH, we consider here the Pati--Salam framework. We show that the Fermi scale is radiatively induced while the GUT scale is held fixed to a phenomenologically viable value. 
Explicitly, we first extend the colour group to an $\SU(4)_{\mathrm{PS}}$ of leptocolour.  Differently from the original Pati--Salam construction, we extend the global symmetry of the Higgs sector to be $\SU(4)_{\chi}$ rather than  $\SUL\times \SUR$.  We indicate the full non-abelian structure by $G=\SU(4)_{\chi}\times\SU(4)_{\mathrm{PS}}$. For simplicity, we consider only one generation of fermions and gauge the EW $\SUL\times{\mathrm{U}}(1)_Y$ subgroup of $\SU(4)_{\chi}$. In the original work  \cite{Pati:1974yy}, the full $\SUL\times \SUR$ was gauged; the consequences of
adapting this to the present model will be considered elsewhere.
   
We want to identify a minimal number of scalar degrees of freedom to achieve the desired symmetry breaking pattern and the radiative generation of the Fermi scale. We find this set to consist of two scalar multiplets with the following assignments with respect to $G=\SU(4)_{\chi}\times\SU(4)_{\mathrm{PS}}$:
	\begin{equation}
	M\sim (6,1), \qquad P\sim (1,15)\footnote{We use the adjoint representation to be  able to preserve the 
	    $\mathrm{U}(1)_{\mathrm{B-L}}$ subgroup of $\SU(4)_{\mathrm{PS}}$ in addition to $\SU(3)_{\mathrm{c}}$},
	\end{equation}
	and thus  the scalars transform under $G$ as 
	\begin{equation}
	    \label{eq:}
	M\rightarrow gMg^T\quad\text{and}\quad P\rightarrow hPh^{-1},	    
	\end{equation}
	where $g\in\SU(4)_{\chi}$ 
	and $h\in\SU(4)_{\mathrm{PS}}$. 
	
	The most general renormalizable scalar potential then reads:
	\begin{equation}
	    \label{eq:Vtot}
	    V=V_M+V_P+V_{MP},
	\end{equation}
	where	
	\begin{equation}
	    \label{eq:pot}
	    \begin{split}
		V_M=&\frac{1}{2}m_M^2\mathrm{Tr}[M^{\dagger} M]
		+\frac{\lambda_M}{4}\mathrm{Tr}[M^{\dagger}M]^2, \\
	    V_P=&m_P^2\mathrm{Tr}[P^{2}]+\lambda_{P1}\mathrm{Tr}[P^2]^2\\
	    &+\lambda_{P2}\mathrm{Tr}[P^4], \\
	    V_{MP}=&\frac{\lambda_{MP}}{2}\mathrm{Tr}[M^{\dagger}M]\mathrm{Tr}[P^2].
	    \end{split}
	\end{equation}

	We parameterize $P$ as
	\begin{equation}
	    \label{eq:}
	    P=p_aT^a,    
	\end{equation}
	and $M$ as
	\begin{equation}
	    \label{eq:M}
	    M=\left[\frac{\sigma}{2}+\ii\, \sqrt{2}\, \Pi_i X ^i\right]E ,
	\end{equation}
	where  the index $a$ runs over the 15 generators of $\SU(4)_{\mathrm{PS}}$ and $i$ over the 5 broken generators $X^i$ of $SU(4)_\chi$, 
	and   $E$ is a fully antisymmetric matrix. The vacuum expectation value (vev) of $M$ is then given by  $\langle M\rangle=\frac{v_0}{2}E$.  
	
	We have one more scalar that is required to acquire a vev,  $\langle P\rangle=b_0 T^{15}$, to break the leptocolour group to 
	$\SU(3)_{\mathrm{c}}\times \mathrm{U}(1)_{\mathrm{B}-\mathrm{L}}$. 
	Minimising the tree-level potential we obtain:
	\begin{equation}
	    \label{eq:vb0}
	    \begin{split}
		b_0^2=&\frac{6\lambda_{MP}m_M^2-12\lambda_M m_P^2}{\lambda_M(12\lambda_{P1}+7\lambda_{P2})-3\lambda_{MP}^2},\\
		v_0^2=&\frac{6\lambda_{MP}m_P^2-(12\lambda_{P1}+7\lambda_{P2})m_M^2}
		    {\lambda_M(12\lambda_{P1}+7\lambda_{P2})-3\lambda_{MP}^2}.
	    \end{split}
	\end{equation}
	These coupled expressions for the vevs of the two scalar fields would require couplings that are vastly different in value to be able to accommodate simultaneously the GUT and the Fermi scale. We will show in the next section that the Fermi scale can, de facto, be radiatively generated. 
  
    \section{Radiative Fermi scale}
    The symmetry breaking pattern $\SU(4)\rightarrow\Sp(4)$ has been studied at length in literature \cite{Appelquist:1999dq, Cacciapaglia:2014uja}. 
The EW gauge group can be embedded in $\SU(4)$ in different ways with respect to the vacuum.  We parameterise this freedom  by an angle $\theta$. The matrix $E$ in~\eqref{eq:M} is correspondingly replaced by $E_{\theta}$,
	\begin{equation}
	E_{\theta}= \sin\theta 
	\begin{pmatrix} 
	0 & 1\\
	-1& 0\\
	\end{pmatrix}
+\cos\theta 
\begin{pmatrix} \ii \, \sigma_2 & 0\\
0 & -\ii \, \sigma_2 \\
	\end{pmatrix}.
	\end{equation}

For $\theta=0$, the EW symmetry remains unbroken and for $\theta=\pi/2$ the EW symmetry directly breaks to $\mathrm{U}(1)_Q$. The specific value of $\theta$ must be determined dynamically once the EW and top quark quantum corrections are taken into account.  As shown in \cite{Alanne:2014kea,Gertov:2015xma}, these corrections favour small values of $\theta$, and consequently the Fermi scale, $v_{w} = v_0 \sin \theta$, lies well below the  spontaneous symmetry breaking scale  $v_0$. Furthermore, the radiative corrections provide a mass for the pGB Higgs via the Coleman--Weinberg mechanism. 
	
Following \cite{Alanne:2014kea}, 
 in the $\overline{\mathrm{MS}}$ scheme, the one-loop potential is
	\begin{equation}
	    \label{eq:deltaV}
	    ~\hspace{-0.35cm}\delta V=\frac{1}{64\pi^2}\mathrm{Str}\left[{\cal M}^4(\Phi)\left(\log\frac{{\cal M}^2(\Phi)}
		{\mu_0^2}-C\right)\right],
	\end{equation}
	where ${\cal M}(\Phi)$ is the tree-level mass matrix of the scalar fields that we denote collectively as $\Phi$, computed using the background field method.  The supertrace 
	$\mathrm{Str}$, is defined by
	\begin{equation}
	    \mathrm{Str} = \sum_{\text{scalars}}-2\sum_{\text{fermions}}+3\sum_{\text{vectors}}.
	    \end{equation}
	We have $\displaystyle{C=3/2}$ for scalars and fermions, while $C=\displaystyle{5/6}$ for the gauge bosons and we include contributions from the EW gauge bosons and the top quark.  
		
	We fix the renormalization scale by requiring that the vev $v=\langle\sigma\rangle$ is given by the tree-level value $v_0$ while the one for $\langle p_{15}\rangle=b$ is determined by minimising the full one-loop potential along with the dynamical value of $\theta$.   
		
Three states have the quantum numbers of the Higgs, i.e. $\Pi_4$, $ \sigma$ and $p_{15}$. In the numerical analysis we show that a  small value of $\theta$ is radiatively generated. Therefore the observed Higgs is mostly the pGB $\Pi_4$ with a tiny admixture of $\sigma$, and it is constrained, in the minimisation analysis,  to give the observed value of its mass. 
		
\begin{figure}
		\begin{center}
		    \includegraphics[width=0.44\textwidth]{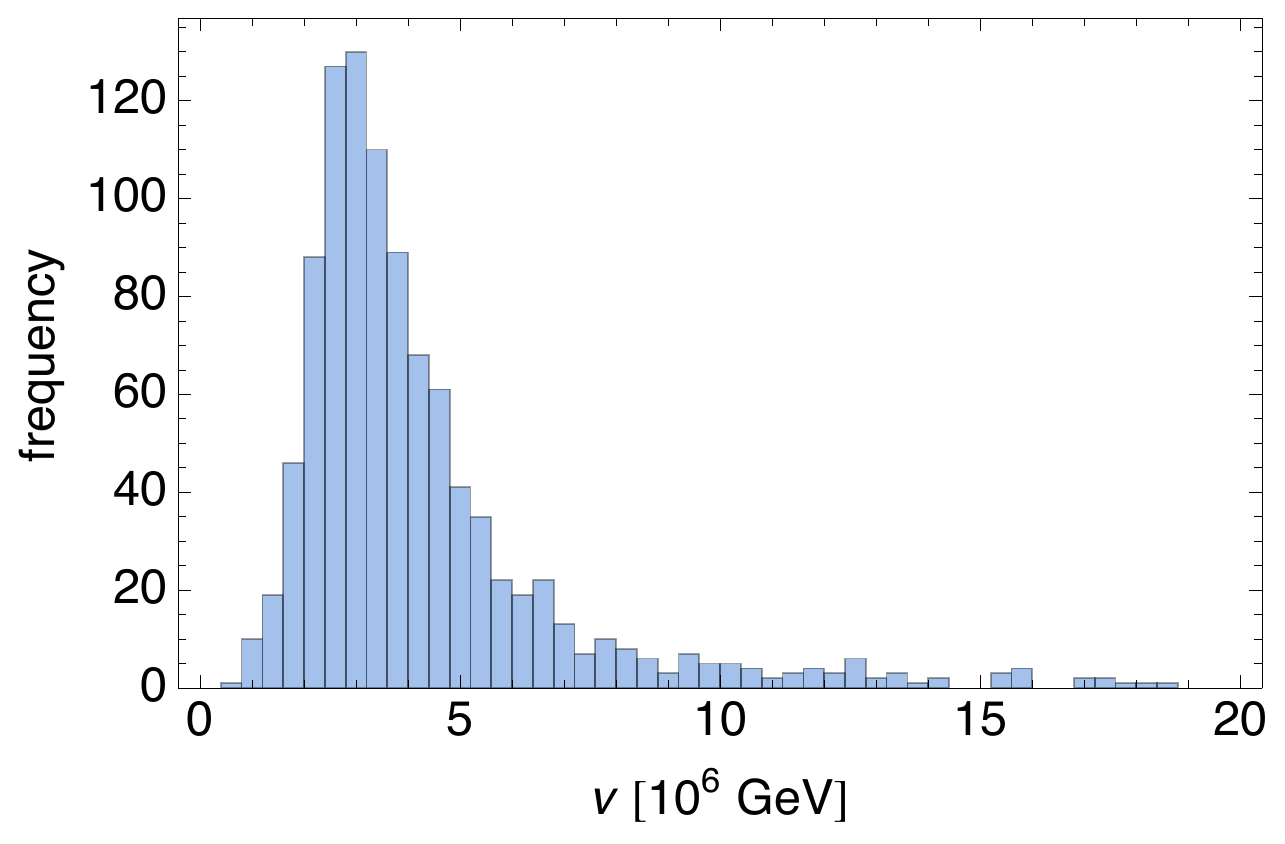}
		\end{center}\vspace{-10pt}
		\caption{Distribution of values of $v$ with $b=2.5\cdot 10^6$~GeV.}
		\label{fig:vL2p5}
	    \end{figure}

\begin{figure}
		\begin{center}
		    \includegraphics[width=0.45\textwidth]{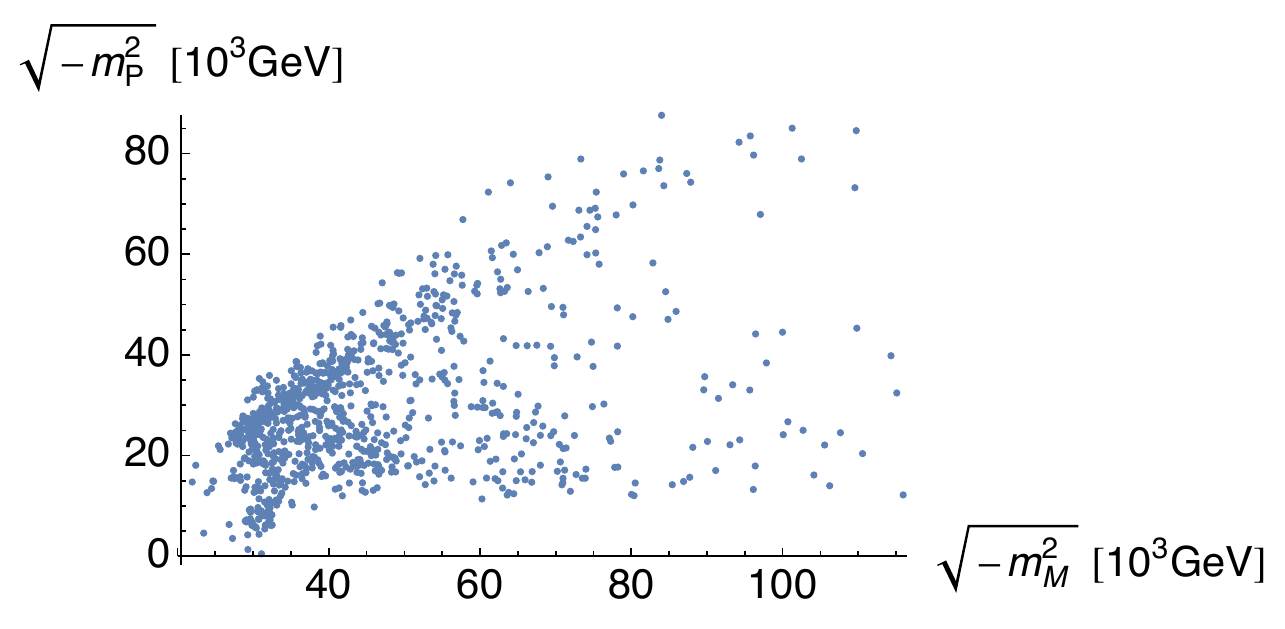}
		\end{center}\vspace{-10pt}
		\caption{Distribution of $\sqrt{-m_M^2}$ vs $\sqrt{-m_P^2}$.}
		\label{fig:mass}
	    \end{figure}

We explore the  parameter space $(b,\,v,\,\theta,\,\lambda_M,\,\lambda_{MP},\,\lambda_{P1},\lambda_{P2}$) by assuming the leptocolour breaking scale to be just above the experimental bound, i.e. $b=2.5\cdot 10^6$~GeV\ \footnote{We have also explored regions of parameter space where $b$ is allowed to be significantly larger, e.g. $b\sim{\cal O}(10^{10}\ \mathrm{GeV})$. We find that it is still possible to have $b\sim v$ with $\vw$ emerging as a radiatively generated scale.}, and we check that the tree-level potential, Eq.~\eqref{eq:Vtot}, is bounded from below. 
In Fig.~\ref{fig:vL2p5} we show the resulting values of the scale $v$ of the global symmetry breaking.
The preferred values of $v$ are roughly of the order of $b$, and this feature is reflected also in the Lagrangian mass parameters as shown in Fig.~\ref{fig:mass}. To produce the correct Fermi scale this implies that small values of the angle $\theta$ are favoured.		    

\begin{figure}
		\begin{center}
		    \includegraphics[width=0.48\textwidth]{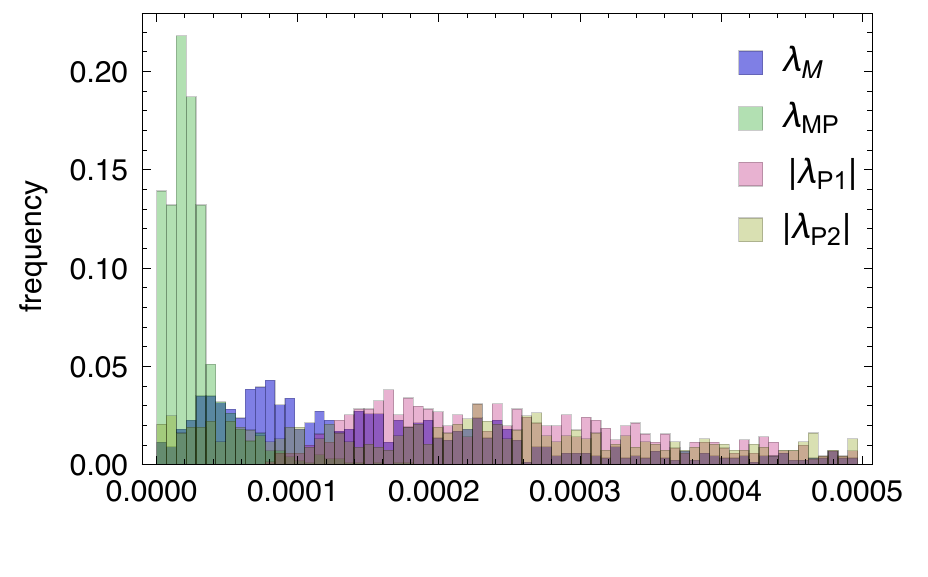}
		\end{center}\vspace{-10pt}
		\caption{Distribution of the quartic scalar couplings of 1000 viable 
		scanned points with $b=2.5\cdot 10^6$~GeV.}
		\label{fig:lambda}
	    \end{figure}

We also find the distribution of quartic couplings shown in Fig.\,\ref{fig:lambda}.  We see that the values of the quartic couplings are overall small, less than $10^{-2}$ for all scanned points; in particular $\lambda_{MP}\sim {\cal O}(10^{-4})$. Generally we can understand this as follows:  The minimisation procedure gives a relation between the couplings of the scalar potential and the vacuum angle $\theta$. In the limit of equal self-couplings\,\footnote{In the general case the minimisation requirement produces a non-linear logarithmic equation which cannot be solved analytically.}
and $v=b$, we find $\lambda\sim\sin^2\theta$.  Furthermore, we fix the mass of the lightest scalar to be 125 GeV. This is sensitive in particular to the coupling $\lambda_{MP}$, explaining the restriction on its values as shown in Fig.\,\ref{fig:lambda}.

 The numerical analysis shows that it is possible to find viable parameter space of the theory allowing to abide all the phenomenological constraints for a successful Pati--Salam extension of the EGH model\footnote{Because of the relatively large GUT scale, smaller values of the scalar couplings are needed reducing the parameter space of the theory. This in not in contradiction with the results of \cite{Gertov:2015xma}  that favoured a lower scale for the vev of the $\sigma$-field and the scalar couplings were allowed to span a much larger parameter space. }. 
    
\section{Conclusions and Outlook}

We have shown that a dynamical mechanism, in the form of radiative corrections, can generate the desired hierarchy between the EW and GUT scales. 

The natural starting point is the EGH model with the symmetry breaking pattern SU(4)$_\chi\rightarrow$Sp(4)$_\chi$. Of the five GBs, three become the longitudinal components of the EW gauge bosons while the fourth one, via mixing with the radial mode, is identified with the observed Higgs particle. The remaining GB, $\Pi_5$, can be a dark matter candidate \cite{Alanne:2014kea}.

The heaviest states in the spectrum are the massive spin-one leptoquarks, whose masses are constrained by experiments to be above $10^3$ TeV. The  heavy scalars have masses ${\cal O}(10-100\,{\rm{TeV}})$. The lightest states of the spectrum are the pGBs. 

In this scenario all scalar self-couplings are generally very small and hence lead to testable consequences.
In practice, measuring the trilinear Higgs coupling at the LHC \cite{Barger:2013jfa} is sufficient to constrain this framework.

 A more in-depth analysis of dark matter can be done following~\cite{Alanne:2014kea}. This would require adding a small $\SU(4)_{\chi}$-breaking mass for $\Pi_5$. 

Another interesting avenue to explore is neutrino masses and mixings. One can, for example, accommodate right-handed neutrinos into the fundamental representation of SU(4)$_\chi$. This would naturally lead to Dirac masses for neutrinos. More generally, various see-saw scenarios can also be realised in this setup and studied in connection with collider phenomenology \cite{Deppisch:2015qwa}. For example, one could gauge the full chiral symmetry subgroup SU(2)$_{\mathrm{L}}\times$SU(2)$_{\mathrm{R}}$ (or even the entire SU(4)$_\chi$). This would allow implementing the type-II see-saw mechanism \cite{Mohapatra:1980yp,Magg:1980ut,Schechter:1980gr}. 

In addition, the scalar sectors could also drive cosmic inflation \cite{Starobinsky:1979ty,Guth:1980zm,Linde:1981mu} known to also prefer small scalar self-couplings. 

It is therefore clear from the above that new prospects are ready to be explored within the  radiatively generated Fermi scale in GUT scenarios. 
\\

\acknowledgments

We thank Borut Bajc for enlightening discussions. The $\mathrm{CP}^3$-Origins centre is partially funded by the Danish National Research Foundation, grant number DNRF90. 
T.A. would like to thank the Finnish Cultural Foundation for financial support.
K.T. acknowledges support from the Academy of Finland, project 267842.

\bibliography{Elementary}

\end{document}